\documentclass[12pt]{article}
\usepackage{dcolumn}
\usepackage{color}
\usepackage{latexsym}
\usepackage{bm}
\begin{document}
\date{}
%%%%%%%%%%%%%%%%%%%%%%%%%%%%%%%%%%%%%%%%%%%%%%%%%%%%%%%%%%%%%%%%%%%%%%%%%%
%\title{\textbf{Covariant Anomalies and Hawking Radiation}}
%\author{{Rabin Banerjee}\thanks{E-mail: rabin@bose.res.in}, \ {Shailesh Kulkarni}\thanks{E-mail: shailesh@bose.res.in}\\
%\\\textit{S.~N.~Bose National Centre for Basic Sciences,}
%\\\textit{JD Block, Sector III, Salt Lake, Kolkata-700098, India}}
%%%%%%%%%%%%%%%%%%%%%%%%%%%%%%%%%%%%%%%%%%%%%%%%%%%%%%%%%%%%%%%%%%%%%%%%%%%%%%
\title{Hawking Radiation, Covariant Boundary Conditions and Vacuum States}
\author{{Rabin Banerjee}\thanks{E-mail: rabin@bose.res.in}, \  {{Shailesh Kulkarni}\thanks{E-mail: shailesh@bose.res.in}}\\
%\textit{Institute of Quantum Science, College of Science and}\\
%\textit{Technology, Nihon University, Tokyo 101-8308, Japan.}\\
\textit{S.N. Bose National Centre for Basic Sciences,}\\
\textit{JD Block, Sector III, Salt Lake, Kolkata-700098, India}}
%%%%%%%%%%%%%%%%%%%%%%%%%%%%%%%%%%%%%%%%%%%%%%%%%%%%%%%%%%%%%%%%
\maketitle
%%%%%%%%%%%%%%%%%%%%%%%%%%%%%%%%%%%%%%%%%%%%%%%%%%%%%%%%%%%%%%%
\begin{quotation}
\noindent \normalsize
The basic characteristics of the covariant chiral current $<J_{\mu}>$ and 
the covariant chiral energy-momentum tensor $<T_{\mu\nu}>$  are obtained from a chiral effective action.
These results are used to justify the covariant boundary condition used
in recent approaches \cite{Isowilczek,Isoumtwilczek,shailesh,shailesh2,Banerjee} of computing the Hawking flux from chiral gauge and gravitational anomalies.
We also discuss a connection of our results with the conventional calculation of nonchiral currents and stress tensors in different 
(Unruh, Hartle-Hawking and Boulware) states.
\end{quotation}

%%%%%%%%%%%%%%%%%%%%%%%%%%%%%%%%%%%%%%%%%%%%%%%%%%%%%%%%%%%%%%%%
\section{Introduction}

The motivation of this paper is to provide a clear understanding of
the covariant boundary condition used in the recent analysis 
\cite{Isowilczek,Isoumtwilczek,shailesh,shailesh2, Banerjee} of deriving the Hawking flux using chiral gauge and gravitational anomalies. Besides this we also reveal certain
new features in chiral currents and energy-momentum tensors which are useful in exhibiting their connection with the standard nonchiral expressions.

Long ago, Hawking \cite{Hawking} proposed an idea that black holes evaporate,
due to quantum particle creation, and behave like thermal bodies with 
an appropriate temperature. This  is essentially a consequence of quantisation of matter in a background spacetime having an event horizon. There are several approaches to derive the Hawking effect \cite{gibbons, parikh, paddy, Fulling}.
Recently, Wilczek and collaborators \cite{Robwilczek, Isowilczek} gave an interesting method to compute
the Hawking fluxes using chiral gauge and  gravitational(diffeomorphism) anomalies. 
It rests on the fact that the effective theory near the event horizon is a two dimensional chiral theory which, therefore, has gauge and gravitational anomalies.
This method is expected to hold in any dimensions.
In this sense it is distinct from the trace anomaly method \cite{Fulling}
which was formulated in two dimensions\footnote{For a connection
 of the trace anomaly method with the diffeomorphism anomaly approach,
 see \cite{bonora1, bonora2}}. However, an unpleasant
feature of \cite{Robwilczek, Isowilczek} was that whereas the 
 expressions for chiral anomalies were taken to be consistent, the boundary conditions required to fix the arbitrary constants were covariant. 
This was rectified by us \cite{shailesh} and a simplified derivation using
only covariant forms was presented. It might be recalled that there are
two types of chiral anomalies - covariant and consistent. Covariant
anomalies transform covariantly under the gauge or general coordinate
transformation but do not satisfy the Wess-Zumino consistency condition.
Consistent anomalies, on the contrary, behave the other way. Covariant
and consistent expressions are related by local counterterms
\cite{Bardeen,rabin2,rabin1, bertlmann, Fujikawa}.

 In another new development also based on chiral gauge and gravitational anomalies, Hawking fluxes were obtained by us \cite{shailesh2, Banerjee}. Contrary to the earlier approaches \cite{Robwilczek,Isowilczek,Isoumtwilczek,shailesh} a
splitting of the space in different regions (near to and away from the horizon) using discontinuous (step)functions was avoided. This split, 
apart from requiring the necessity of both the normal and anomalous
Ward identities, poses certain conceptual issues \cite{singleton}.
In \cite{shailesh2, Banerjee} the only input was the structure of the covariant anomaly while retaining the original covariant boundary condition, i.e the vanishing of the covariant current/energy-momentum (EM) tensor at the event horizon.

It is thus clear that the covariant boundary condition plays an important role in the computation of Hawking fluxes. However, a precise understanding of this boundary condition is still missing. Here we give a detailed analysis for this particular choice of boundary condition.
It turns out that, with this choice of covariant boundary condition,
the components for covariant current/EM tensors ($J^{r}, T^{r}_{\quad t}$)
 obtained from solving the anomaly equation match exactly with the expectation values of the current/EM tensors, obtained from the chiral
effective action, taken by imposing the regularity condition on the outgoing
modes at the future horizon. Furthermore, we discuss the connection of our results with those found by a standard use of boundary conditions on nonchiral 
(anomaly free) currents and EM tensors. Indeed we are able to  show that our results are equivalent to the 
choice of the Unruh vacuum for a nonchiral theory. This choice, it may be recalled, is natural for discussing Hawking flux. 

 In section 2 we provide a generalisation of our recent approach \cite{shailesh2,Banerjee} of computing fluxes. The covariant current/EM tensor following from
a chiral effective action, suitably modified by a local counterterm, 
are obtained in section 3. The role of chirality in imposing constraints
on the structure of the current/EM tensor is elucidated. The arbitrary coefficients in $J_{\mu}, T_{\mu\nu}$  are fixed by imposing appropriate regularity conditions 
on the outgoing modes at the future event horizon (section 4). Here we also discuss the relation of the results obtained for a chiral theory, subjected to the regularity conditions, with those found in a nonchiral theory in different vacua. Some examples are given in section 5. Our concluding  remarks are contained in section 6. Finally, there is an appendix discussing the connection between
the trace anomaly and gravitational anomaly for a ($1+1$) dimensional chiral theory.   
\section{Charge and energy flux from covariant anomaly}
Consider a generic spherically symmetric black hole represented by the metric,
\begin{equation}
ds^2 = f(r)dt^2 -\frac{1}{h(r)}dr^2 -r^2(d\theta^2 + sin^2\theta d\phi^2) \label{1}
\end{equation}
where $f(r)$ and $h(r)$ are the metric coefficients. The event horizon is
defined by $f(r_{h}) = h(r_{h}) =0$. Also, in the asymptotic limit the metric (\ref{1}) become Minkowskian i.e $f(r\rightarrow \infty)=h(r\rightarrow \infty) = 1$ and $f''(r\rightarrow \infty) = f'''(r\rightarrow \infty)= h''(r\rightarrow \infty)= h'''(r\rightarrow \infty)= 0$. \\
Now consider quantum fields (scalar or fermionic)propagating on this
background. It was shown that \cite{Robwilczek,Isowilczek}, by using a dimensional reduction technique, the effective field theory near the event horizon becomes two dimensional with the metric given by the  $r-t$ section of (\ref{1})
\begin{equation}
ds^2=f(r)dt^2 -\frac{1}{h(r)}dr^2 ~. \label{2}
\end{equation}
Note that $\sqrt{-g} = \sqrt{-det g_{\mu\nu}} = \sqrt{\frac{f}{h}} \ne1$
(unless $f(r) = h(r)$). On this two dimensional background, the modes
which are going in to the black hole (for example left moving modes) are lost and the effective theory become chiral. Two dimensional chiral theory possesses gravitational anomaly and, if gauge fields are present, also gauge anomaly \cite{Kolprath, Witten, Bardeen, rabin2, rabin1, bertlmann, Fujikawa}. Hawking radiation, which is necessary to  cancel these  anomalies, were obtained by solving the anomalous Ward identity near the horizon and  the usual (i.e anomaly free) conservation equations which are valid far away from the horizon\cite{Robwilczek,Isowilczek}. This approach used consistent forms for gauge and gravitational anomaly. However, the boundary condition used to fix the arbitrary constants was covariant. As already  stated, a reformulation of this approach using only covariant structures was given by us \cite{shailesh, shailesh2}.\\
An efficient  and economical way to obtain the Hawking flux was discussed in \cite{Banerjee} where the computation involved only the expressions for      
 anomalous covariant Ward identities and the covariant boundary conditions. 
 The splitting of space into two regions \cite{Robwilczek, Isowilczek,Isoumtwilczek, shailesh} is avoided. Here we would first  generalise this new approach for the generic black hole (\ref{2}). This would also help in setting up the conventions and 
 introduce certain equations that are essential for the subsequent 
 analysis.

  As already stated the effective theory near the event horizon is a two dimensional
chiral theory. The relevant contribution comes from the outgoing (right moving)
 modes only. For these modes the expression for covariant gauge anomaly
is given by \cite{bertlmann,Fujikawa},
\begin{equation}
\nabla_{\mu}J^{\mu}= -\frac{e^2}{4\pi\sqrt{-g}}\epsilon^{\alpha\beta}F_{\alpha\beta}\label{3} 
\end{equation}
$\epsilon^{01}=-\epsilon^{10} =1$,  $F_{\mu\nu} = \partial_{\mu}A_{\nu} -\partial_{\nu}A_{\mu}$ and the gauge potential is defined as $A_{t} = -\frac{Q}{r}$. For a static background, the above equation becomes,
\begin{equation}
\partial_{r}(\sqrt{-g}J^{r}) = \frac{e^{2}}{2\pi}\partial_{r}A_{t}.\label{3.1}
\end{equation}
Solving this equation we get 
\begin{equation}
\sqrt{-g} J^{r} = c_{H} + \frac{e^2}{2\pi}[A_{t}(r) -A_{t}(r_{h})].
\end{equation}
Here $c_{H}$ is an integration constant which can be fixed by 
imposing the covariant boundary condition i.e  covariant current ($J^{r}$)
must vanish at the event horizon, 
\begin{equation}
J^{r}(r=r_{h}) = 0~. \label{a}
\end{equation}
 Hence we get $c_{H} = 0$ and the expression for the current becomes,
\begin{equation}
J^{r} = \frac{e^{2}}{2\pi\sqrt{-g}}[A_{t}(r) -A_{t}(r_{h})]. \label{4}
\end{equation}      
Note that the Hawking flux is measured at infinity where there is no 
anomaly. This necessitated a split of space into two distinct regions -
one near the horizon and one away from it - and the use of two Ward
identities \cite{Robwilczek,Isowilczek,Isoumtwilczek,shailesh, shailesh2}. This is redundant if we observe that the anomaly (\ref{3.1})
vanishes at the asymptotic infinity. Consequently, in this approach,
the flux is directly obtained from the asymptotic infinity limit of 
(\ref{4}):  
\begin{equation}
Charge flux = J^{r}(r\rightarrow \infty) = -\frac{e^2 A_{t}(r_{h})}{2\pi} =  \frac{e^2 Q}{2\pi r_{h}}. \label{5}
\end{equation}
This reproduces the familiar expression for the charge flux \cite{Robwilczek, Isowilczek, shailesh, shailesh2, Banerjee} .\\
Next, we consider the expression for the two dimensional covariant gravitational Ward identity \cite{Isowilczek, Isoumtwilczek, shailesh, shailesh2},
\begin{equation}
\nabla_{\mu}T^{\mu\nu} = J_{\mu}F^{\mu\nu} +
\frac{\epsilon^{\nu\mu}}{96\pi\sqrt{-g}}\nabla_{\mu}R \label{6}
\end{equation} 
where the first term is the classical contribution (Lorentz force) and 
the second is the covariant gravitational anomaly \cite{Kolprath,Witten, Fulling3}.
Here $R$ is the Ricci scalar and for the metric (\ref{2}) it is given by
\begin{equation}
R = \frac{f'' h}{f} + \frac{f'h'}{2f} - \frac{f'^{2}h}{2f^2}. \label{ricci}
\end{equation}
By simplifying (\ref{6})we get, in the static background, 
\begin{equation}
\partial_{r}(\sqrt{-g}T^{r}_{\quad t}) = \partial_{r}N^{r}_{t}(r) 
 -\frac{e^2 A_{t}(r_{h})}{2\pi} \partial_{r}A_{t}(r) + \partial_{r}(\frac{e^2 A^{2}_{t}(r)}{4\pi})\label{7}
\end{equation}
where
\begin{equation}
N^{r}_{t} = \frac{1}{96\pi}\left( hf'' + \frac{f'h'}{2} - \frac{f'^{2}h}{f}\right). \label{nrt}
\end{equation}
The solution for (\ref{7}) is given by
\begin{equation}
\sqrt{-g}T^{r}_{\quad t} = b_{H} + [N^{r}_{t}(r) - N^{r}_{t}(r_{h})]
 + \frac{e^{2}A_{t}^{2}(r_{h})}{4\pi} - \frac{e^2}{2\pi}A_{t}(r_{h})A_{t}(r) 
 + \frac{e^2 A_{t}^{2}(r)}{4\pi} ~. \label{8}
\end{equation}
Here $b_{H}$ is an integration constant . Implementing the covariant boundary condition, namely, the vanishing of cova‌riant EM tensor at the event
horizon, 
\begin{equation}
T^{r}_{\quad t}(r=r_{h}) = 0 \label{b}
\end{equation}
yields $b_{H} = 0$. Hence (\ref{8}) reads
\begin{equation}
\sqrt{-g}T^{r}_{\quad t}(r) = [N^{r}_{t}(r) - N^{r}_{t}(r_{h})]
 + \frac{e^2}{4\pi}[A_{t}(r) -A_{t}(r_{h})]^2 ~. \label{9}
\end{equation} 
Since the covariant gravitational anomaly vanishes asymptotically,
we can compute the energy flux as before by taking the asymptotic limit of (\ref{9})
\begin{equation}
energy flux = T^{r}_{\quad t}(r\rightarrow \infty) = -N^{r}_{t}(r_{h}) + \frac{e^2A_{t}^2(r_{h})}{4\pi} = \frac{1}{192\pi}f'(r_{h})g'(r_{h}) + \frac{e^2 Q^2}{4\pi r_{h}^2}~.\label{10}
\end{equation}  
This reproduces the expression for the Hawking flux found by using the anomaly cancelling approach of \cite{Isowilczek,Isoumtwilczek,shailesh, shailesh2}.\\
It is now clear that the covariant boundary conditions play a crucial
role in the computation of Hawking fluxes using chiral gauge and gravitational anomalies, either in the approach based on the anomaly
cancelling mechanism 
\cite{Isowilczek,Isoumtwilczek,shailesh,shailesh2} or in the more direct approach \cite{Banerjee} reviewed here. Therefore it is worthwhile
to study it in some detail. We adopt the following strategy. The expressions
for the expectation values of the covariant current and EM tensor will
be deduced from the chiral effective action, suitably modified by a local
counterterm. Local structures are obtained by introducing auxiliary
variables whose solutions contain arbitrary constants. These constants 
are fixed by imposing regularity conditions on the outgoing modes at the future event horizon. The final results are found to match exactly with the corresponding 
expressions for the covariant current (\ref{4}) and EM tensor (\ref{9}), which were derived by using the covariant boundary conditions (\ref{a},\ref{b}). Subsequently we show that our results are consistent with the imposition of the Unruh vacuum on usual (nonchiral)
expressions.

\section{Covariant current and EM tensor from chiral effective action}

The two dimensional chiral effective action \cite{shailesh2, Leut} is defined
as,  
\begin{equation}
\Gamma_{(H)}= -\frac{1}{3} z(\omega) + z(A) \label{11}
\end{equation}
where $A_{\mu}$ and $\omega_{\mu}$ are the gauge field and the spin connection, respectively, and,
\begin{equation}
z(v) = \frac{1}{4\pi}\int d^2x d^2y \epsilon^{\mu\nu}\partial_\mu v_\nu(x) \Delta^{-1}(x, y)
\partial_\rho[(\epsilon^{\rho\sigma} + \sqrt{-g}g^{\rho\sigma})v_\sigma(y)]
\label{12}
\end{equation}
Here $\Delta^{-1}$ is the inverse of d'Alembertian $\Delta=\nabla^{\mu}\nabla_{\mu}  = \frac{1}{\sqrt{-g}}\partial_{\mu}(\sqrt{-g}g^{\mu\nu}\partial_{\nu})$. 
From a variation of this effective action the energy momentum tensor and the gauge current are computed. These are shown in the literature \cite{Kolprath, Witten, Bardeen, rabin2, rabin1, bertlmann, Fujikawa} as consistent forms. To get their covariant forms in which we are interested, however, appropriate local polynomials have to be added. This is possible since energy momentum tensors and currents are only defined modulo local polynomials. We obtain,
\begin{equation}
\delta \Gamma_{H} = \int d^2x  \sqrt{-g}\left( \frac{1}{2}\delta g_{\mu\nu} T^{\mu\nu} + \delta A_{\mu}J^{\mu}\right) 
+  l \label{13} 
\end{equation}
where the local polynomial is given by \cite{Leut},
\begin{equation}
 l = \frac{1}{4\pi}\int d^{2}x \ \epsilon^{\mu\nu}(A_{\mu}\delta A_{\nu} - 
 \frac{1}{3}w_{\mu}\delta w_{\nu} - \frac{1}{24}R e_{\mu}^{a}\delta e_{\nu}^{a})
\end{equation}
The covariant energy momentum tensor $T^{\mu\nu}$ and the covariant gauge current $J^{\mu}$ are read-off from the above relations as \cite{shailesh2, Leut},
\begin{eqnarray}
T^{\mu}_{\quad \nu} = \frac{e^2}{4\pi}\left(D^{\mu}B D_{\nu}B\right) \nonumber\\
 +\frac{1}{4\pi}\left(\frac{1}{48}D^{\mu}G D_{\nu}G 
-\frac{1}{24} D^{\mu} D_{\nu}G + \frac{1}{24}\delta^{\mu}_{\nu}R\right)
\label{14} 
\end{eqnarray}
 \begin{equation}
J^{\mu} = -\frac{e^2}{2\pi}D^{\mu}B.\label{15}
\end{equation}
Note the presence of the chiral covariant derivative $D_{\mu}$ expressed
in terms of the usual covariant derivative $\nabla_{\mu}$,
\begin{equation}
D_{\mu} = \nabla_{\mu} - \bar\epsilon_{\mu\nu}\nabla^{\nu} = -\bar\epsilon_{\mu\nu}D^{\nu}, \label{16}
\end{equation} 
where $\bar\epsilon_{\mu\nu}=\sqrt{-g}\epsilon_{\mu\nu}$
and $\bar\epsilon^{\mu\nu}=\frac{1}{\sqrt{-g}}\epsilon^{\mu\nu}$.
The auxiliary fields $B$ and $G$ in (\ref{14},\ref{15}) are defined as
\begin{eqnarray}
B(x) &=& \int \ d^2 y \ \sqrt{-g} \Delta^{-1}(x,y)\bar\epsilon^{\mu\nu}
\partial_{\mu}A_{\nu}(y) \label{17}\\
G(x) &=& \int \ d^2 y \ \sqrt{-g}\Delta^{-1}(x,y)R(y) \label{18}
\end{eqnarray}
so that they satisfy 
\begin{eqnarray}
\Delta B(x) &=& \bar\epsilon^{\mu\nu}\partial_{\mu}A_{\nu}(x)\label{19}\\
\Delta G(x)&=& R(x)\label{20}
\end{eqnarray}    
where $R$ is given by (\ref{ricci}).\\
As a simple consistency check the covariant Ward identities (\ref{3},\ref{6})
 are obtained from (\ref{15},\ref{14}). For example, using (\ref{15}), (\ref{17}) and (\ref{19}), we find, 
\begin{equation}
\nabla_{\mu}J^{\mu}= -\frac{e^2}{2\pi}\Delta B = -\frac{e^2}{2\pi}\bar \epsilon^{\mu\nu}\partial_{\mu}A_{\nu} = \frac{-e^{2}}{4\pi\sqrt{-g}}\epsilon^{\mu\nu}F_{\mu\nu}   \label{c}
\end{equation}
reproducing (\ref{3}). Note also the existence of the covariant trace
anomaly\footnote{Observe that the chiral theory has both a 
diffeomorphism anomaly (\ref{6}) and a trace anomaly (\ref{d}). This is distinct from the vector case where there is only a trace anomaly $T^{\mu}_{\quad \mu} = \frac{R}{24\pi}$. No diffeomorphism anomaly exists. See the appendix for more details.} 
following from (\ref{14}),
\begin{equation}
T^{\mu}_{\quad \mu} = \frac{R}{48\pi}~.\label{d}
\end{equation}
The chiral nature of the current (\ref{15}) and the stress tensor 
(\ref{14}) are revealed by the following conditions,
\begin{eqnarray}
J_{\mu} &=& - \bar \epsilon_{\mu\nu}J^{\nu} \label{W1}\\
T_{\mu\nu} &=& -\frac{1}{2}(\bar \epsilon_{\mu\rho}T^{\rho}_{\quad \nu} + \bar \epsilon_{\nu\rho}T^{\rho}_{\quad \mu})  + \frac{g_{\mu\nu}}{2}T^{\alpha}_{\quad \alpha}~. \label{W2}
\end{eqnarray}
which are a consequence of the presence of the chiral derivative (\ref{16}).
This may be compared with the definitions of $J_{\mu}$ and $T_{\mu\nu}$,
 obtained from a Polyakov type action valid for a vector theory, which
do not satisfy the chiral properties (\ref{W1},\ref{W2}). These properties
constrain the structure of $J_{\mu}, T_{\mu\nu}$.\\ 

After solving (\ref{19}) and (\ref{20}) we get, 
\begin{equation}
B(x) = B_{o}(r) -at +b  \  ; \ \partial_{r}B_{o} = \frac{A_{t}(r)+c}{\sqrt{fh}} \label{21} 
\end{equation}
and 
\begin{equation}
G = G_{o}(r) - 4pt + q  \ ; \  \partial_{r}G_{o} = -\frac{1}{\sqrt{fh}}(\frac{f'}{\sqrt{-g}} + z) \label{22} 
\end{equation}
where $a,b,c,p,q$ and $z$ are constants.
Now, by substituting (\ref{21}) in (\ref{15}) we obtain,
\begin{eqnarray}
J^{r}(r)&=& \frac{e^2}{2\pi\sqrt{-g}}[A_{t}(r) + c + a] \label{23}\\
J^{t}(r) &=& \frac{e^2}{2\pi f} [A_{t}(r) + c + a] = \frac{\sqrt{-g}}{f}J^{r}. \label{24}  
\end{eqnarray}
Observe that there is only one independent component of $J_{\mu}$ which
is a consequence of (\ref{W1}). Likewise, by using (\ref{21},\ref{22}) in (\ref{14}) we find
%%%%%%%%%%%%%%%%%%%%%%%%%%%%%%%%%%%%%%%%%%%%%%%%%%%%%%%%%%%%%%%%%%%%%%%%%%%%%%%%%%%%%%%%%%%%%%%%%%%%%%%%%%%%%%%%%%%%%%%%%%%%%%%%%%%%%%%%%%%%%%%%
%\begin{eqnarray}
%T^{r}_{\quad t} &=& \frac{e^2}{4\pi \sqrt{-g}}[A_{t}(r)+c+a]^2 + \frac{1}{12\pi\sqrt{-g}}\left(p - \frac{1}{4}(\frac{f'}{\sqrt{-g}} + z)\right)^2\nonumber\\
%&& + \frac{1}{24\pi \sqrt{-g}}\left[\frac{f'}{\sqrt{-g}}\left(p - \frac{1}{4}(\frac{f'}{\sqrt{-g}} + z)\right) + \frac{1}{4}hf'' - \frac{f'}{8}(\frac{hf'}{f} - h')\right]\label{25}\\
%T^{r}_{\quad r} &=& \frac{-e^2}{4\pi f}[A_{t}(r)+c+a]^2 - \frac{1}{12\pi f}\left(p - \frac{1}{4}(\frac{f'}{\sqrt{-g}} + z)\right)^2\nonumber\\
%&&-\frac{1}{24\pi\sqrt{-g}f}\left[f'\left(p - \frac{1}{4}(\frac{f'}{\sqrt{-g}} + z)\right) + \frac{\sqrt{-g}}{4}hf'' - \frac{\sqrt{-g}f'}{8}(\frac{hf'}{f} - h')\right] + \frac{R}{96\pi}\label{26}\\
%T^{t}_{\quad t} &=&\frac{e^2}{4\pi f}[A_{t}(r)+c+a]^2 - \frac{1}{12\pi f}\left(p - \frac{1}{4}(\frac{f'}{\sqrt{-g}} + z)\right)^2\nonumber\\
%&&+\frac{1}{24\pi \sqrt{-g}f}\left[f'\left(p - \frac{1}{4}(\frac{f'}{\sqrt{-g}} + z)\right) + \frac{\sqrt{-g}}{4}hf'' - \frac{\sqrt{-g}f'}{8}(\frac{hf'}{f} - h')\right] + \frac{R}{96\pi}\label{26'} 
%\end{eqnarray}
%%%%%%%%%%%%%%%%%%%%%%%%%%%%%%%%%%%%%%%%%%%%%%%%%%%%%%%%%%%%%%%%%%%%%%%%%%%%%%%%%%%%%%%%%%%%%%%%%%%%%%%%%%%%%%%%%%%%%%%%%%%%%%%%%%%%%%%%%%%%%%%%
\begin{eqnarray}
T^{r}_{\quad t} &=& \frac{e^2}{4\pi \sqrt{-g}}\bar A^{2}_{t}(r) + \frac{1}{12\pi\sqrt{-g}}\bar P^{2}(r) + 
\frac{1}{24\pi \sqrt{-g}}[\frac{f'}{\sqrt{-g}} \bar P(r) + \bar Q(r)]\label{25}\\
T^{r}_{\quad r} &=& \frac{R}{96\pi} - \frac{\sqrt{-g}}{f}T^{r}_{\quad t} \label{26}\\
T^{t}_{\quad t}&=& -T^{r}_{\quad r}+ \frac{R}{48\pi}\label{26'}
\end{eqnarray}
%%%%%%%%%%%%%%%%%%%%%%%%%%%%%%%%%%%%%%%%%%%%%%%%%%%%%%%%%%%%%%%%%%%%%%%%%%%%%%%%%%%%%%%%%%%%%%%%%%%%%%%%%%%%%%%%%%%%%%%%%%%%%%%%%%%%%%%%%%%%%%%%%%
with $\bar A_{t}(r),\bar P(r)$ and $\bar Q(r)$  defined as
\begin{eqnarray}
\bar A_{t}(r) &=& A_{t}(r)+c+a\label{A}\\
\bar P(r) &=& p - \frac{1}{4}(\frac{f'}{\sqrt{-g}} + z)\label{P}\\
\bar Q(r) &=& \frac{1}{4}hf'' - \frac{f'}{8}(\frac{hf'}{f} - h')~.\label{Q}
\end{eqnarray}
Relation (\ref{26'}) is a consequence of the trace anomaly (\ref{d}) while
(\ref{26}) follows from the chirality criterion (\ref{W2}).
The $r-t$ component of the EM tensor (\ref{25}) calculated above is same as the one given in \cite{sunandan}.\\

To further illuminate the chiral nature, we transform the various components of current/EM tensor to 
null coordinates given by
\begin{eqnarray}
v &=& t + r*  \ ; \  \frac{dr}{dr*} = \sqrt{fh} \label{27}\\
u &=& t -r* \label{28}  
\end{eqnarray}
The metric (\ref{2}) in these coordinates looks like
\begin{equation}
ds^2 = \frac{f(r)}{2}(dudv+dvdu). \label{nullmetric} 
\end{equation}
Finally, the expressions for the current and EM tensors in these coordinates
are given by,
\begin{eqnarray}
J_{u} &=& \frac{1}{2}[J_{t} - \sqrt{fh}J_{r}] =\frac{e^2}{2\pi} \bar A_{t}(r)\label{29}\\
J_{v} &=&  \frac{1}{2}[J_{t} + \sqrt{fh}J_{r}]  = 0 \label{30}
\end{eqnarray}
and  
\begin{eqnarray}
T_{uu} &=& \frac{1}{4}[fT^{t}_{\quad t} - fT^{r}_{\quad r} + 2\sqrt{-g}T^{r}_{\quad t}]\nonumber\\
&=& \frac{e^2}{4\pi }\bar A^{2}_{t}(r) + \frac{1}{12\pi}\bar P^2(r) + \frac{1}{24\pi }\left[\frac{f'}{\sqrt{-g}}\bar P(r) + \bar Q(r)\right] \label{31}\\
T_{uv} &=& \frac{f}{4}[T^{t}_{\quad t} + T^{r}_{\quad r}] = \frac{1}{192\pi}fR \label{32}\\
T_{vv} &=&  \frac{1}{4}[fT^{t}_{\quad t} - fT^{r}_{\quad r} - 2\sqrt{-g}T^{r}_{\quad t}]=0 ~.\label{32'}
\end{eqnarray}
where  extensive use has been made of (\ref{23})till (\ref{26'}).
We now observe that, due to the chiral property, the $J_{v}$ and $T_{vv}$
components vanish everywhere. These correspond to the ingoing modes and are compatible
with the fact, stated earlier, that the near horizon theory is a two
dimensional chiral theory where the ingoing modes are lost. Also, the
structure of $T_{uv}$ is fixed by the trace anomaly.
Only the $J_{u}$ and $T_{uu}$ components involve the undetermined constants.
These will now be determined by considering various vacuum states.
%%%%%%%%%%%%%%%%%%%%%%%%%%%%%%%%%%%%%%%%%%%%%%%%%%%%%%%%%%%%%%%%%%%%%%%%%%%%%%%%
\section{Vacuum states}

In a generic spacetime three different quantum states (vacua) \cite{Fulling2}
are defined by appropriately choosing 'in' and 'out' modes. This general picture is modified  when dealing with a chiral theory since, as shown before, the 'in' modes always vanish. Consequently this leads to a simplification and conditions are imposed only on the 'out' modes. Moreover, these conditions have to be imposed on the horizon
since the chiral theory is valid only there. The natural condition, leading to the 
occurence of Hawking flux, is that a freely falling observer must see a finite amount
of flux at the horizon. This implies that the current (EM tensor) in Kruskal 
coordinates must be regular at the future horizon. Effectively, this is the same condition on the 'out' modes in either the Unruh vacuum \cite{unruh} or the Hartle-Hawking  vacuum \cite{Hartle}. As far as our analysis is concerned this is sufficient to completely determine the form
of $J_{\mu}$ or $T_{\mu\nu}$. We show that their structures are identical to those obtained in the previous section using the covariant boundary condition. 

 A more direct comparison with the conventional results obtained from Unruh or
Hartle-Hawking states is possible. In that case one has to consider the nonchiral expressions \cite{Isoumtwilczek,fabbri1} containing both 'in' and 'out' modes.
We show that, at asymptotic infinity where the flux is measured, our expressions
agree with that calculated from Unruh vacuum only. We discuss this in some detail.

\subsection{Regularity conditions, Unruh and Hartle-Hawking vacua}

In Kruskal coordinates $U$ the current takes the form $J_{U} = - \frac{J_{u}}{\kappa U}$, where $\kappa$ is the surface gravity. Since $J_{U}$ is 
required to be finite at the future horizon where $U\rightarrow \sqrt{r-r_{h}}$ $(r\rightarrow r_{h})$, $J_{u}$ must vanish at $r\rightarrow r_{h}$.
Hence from (\ref{29}) and (\ref{A}) we have,
\begin{equation}
c + a = -A_{t}(r_{h})~. \label{33}
\end{equation}     
Similarly, imposing the condition that $T_{UU} = (\frac{1}{\kappa U})^2 T_{uu}$
must be finite at future horizon leads to $T_{uu}(r\rightarrow r_{h}) =0$. This yields, from (\ref{31}) and (\ref{A}-\ref{Q}),
\begin{equation}
p = \frac{1}{4}(z \pm \sqrt{f'(r_{h})h'(r_{h})})~. \label{34} 
\end{equation}
Using (\ref{33}) and (\ref{34}) in equations (\ref{23}-\ref{26'}) we obtain
the final expressions,
\begin{eqnarray}
J^{r}(r)&=& \frac{e^2}{2\pi\sqrt{-g}}[A_{t}(r) - A_{t}(r_{h})] \label{35}\\
J^{t}(r) &=& \frac{\sqrt{-g}}{f}J^{r}(r) \label{36}
\end{eqnarray}
for the current and the EM tensor,
\begin{eqnarray}
\sqrt{-g}T^{r}_{\quad t} &=& \frac{e^2}{4\pi}[A_{t}(r)-A_{t}(r_{h})]^2 + 
 [N^{r}_{t}(r)-N^{r}_{t}(r_{h})] \label{37}
\end{eqnarray}
while,
%\begin{eqnarray}
%T^{r}_{\quad r} &=& \frac{-e^2}{4\pi f}[A_{t}(r)-A_{t}(r_{h})]^2 -\frac{1}{f}[N^{r}_{t}(r)-N^{r}_{t}(r_{h})] + \frac{R}{96\pi}\label{38}\\
%T^{t}_{\quad t} &=&\frac{e^2}{4\pi f}[A_{t}(r) -A_{t}(r_{h})]^2 +\frac{1}{f}[N^{r}_{t}(r)-N^{r}_{t}(r_{h})] + \frac{R}{96\pi}\label{39}   
%\end{eqnarray}
 $T^{r}_{\quad r}$ and $T^{t}_{\quad t}$ follow from (\ref{26},\ref{26'})
and  $N^{r}_{t}$ is given by (\ref{nrt}).

The expressions for $J^{r}$ (\ref{35}) and $T^{r}_{\quad t}$ (\ref{37})
agree with the corresponding ones given in (\ref{4}) and (\ref{9}).
This shows that the structures for the universal components
$J^{r}, T^{r}_{\quad t}$ obtained by solving the anomalous Ward
identities (\ref{3},\ref{6}) subjected to the covariant boundary conditions
(\ref{a},\ref{b}) exactly coincide with the results computed by demanding
regularity at the future event horizon. 

  It is possible to compare our findings with conventional (nonchiral) computations
where the Hawking flux is obtained in the Unruh vacuum. We begin by considering the conservation equations
for a nonchiral theory that is valid away from the horizon.
Such equations were earlier used in \cite{Robwilczek,Isowilczek,Isoumtwilczek,shailesh}. Conservation
of the gauge current yields \footnote{We use a tilde ($J^{\mu}$) to distinguish nonchiral expressions from chiral ones.},
\begin{equation}
\nabla_{\mu}\tilde J^{\mu} = \frac{1}{\sqrt{-g}}\partial_{\mu}(\sqrt{-g}\tilde J^{\mu}) = 0 \label{new1}
\end{equation}
which, in a static background, leads to,
\begin{equation}
\tilde J^{r} = \frac{C_{1}}{\sqrt{-g}} \label{new2}
\end{equation}
where $C_{1}$ is some constant.

   As is well know there is no regularisation that simultaneously preserves the vector as well as axial vector
gauge invariance. Indeed a vector gauge invariant regularisation resulting in (\ref{new1}) yields the following
axial anomaly,
\begin{equation}
\nabla_{\mu}\tilde J^{5\mu} = \frac{e^2}{2\pi \sqrt{-g}}\epsilon^{\mu\nu}F_{\mu\nu} \ ; \ \tilde J^{5\mu} = \frac{1}{\sqrt{-g}}\epsilon^{\mu\nu}\tilde J_{\nu}~. \label{new3}
\end{equation}
The solution of this Ward identity is given by,
\begin{equation}
\tilde J^{t} = -\frac{1}{f}[C_{2} - \frac{e^2}{\pi}A_{t}(r)]
\end{equation} 
where $C_{2}$ is another constant.

    In the null coordinates introduced in 
(\ref{27},\ref{28},\ref{29},\ref{30}) the various components are defined as ,
\begin{eqnarray}
\tilde J_{u} &=& \frac{1}{2}[C_{1} - C_{2} + \frac{e^2}{\pi}A_{t}(r)],\label{new4}\\
\tilde J_{v} &=& -\frac{1}{2}[C_{1} + C_{2} - \frac{e^2}{\pi}A_{t}(r)]~.\label{new5}
\end{eqnarray}
The constants $C_{1}$, $C_{2}$ are now determined by using
appropriate boundary conditions corresponding to first,
the Unruh state, and then, the Hartle-Hawking state. For
the Unruh state $\tilde J_{u}(r\rightarrow r_{h}) = 0$ 
and $\tilde J_{v}(r \rightarrow \infty) = 0$ yield,
\begin{equation}
C_{1} = - C_{2} = -\frac{e^2}{2\pi}A_{t}(r_{h})~,\label{new6}
\end{equation}
so that, reverting back to $(r,t)$ coordinates, we obtain,
\begin{eqnarray}
\tilde J^{r} &=& -\frac{e^2}{2\pi \sqrt{-g}}A_{t}(r_{h}),\label{new7}\\
\tilde J^{t} &=& \frac{e^2}{\pi f}[A_{t}(r) - \frac{1}{2}A_{t}(r_{h})]~.\label{new8}
\end{eqnarray}
The Hawking charge flux, identified with $\tilde J^{r}(r \rightarrow \infty)$, 
reproduces the desired result (\ref{5}). Expectedly, (\ref{new7},\ref{new8}) differ from our relations (\ref{35},\ref{36}) which are valid only near the horizon. However, at
 asymptotic infinity where the Hawking flux is measured,
 both expressions match, i.e
\begin{eqnarray}
\tilde J^{r}(r\rightarrow \infty) &=& J^{r}(r \rightarrow \infty), \label{compare1}\\
\tilde J^{t}(r\rightarrow \infty) &=& J^{t}(r \rightarrow \infty)~.\label{compare2} 
\end{eqnarray}

  All the above considerations follow identically for 
the stress tensor. Now the relevant conservation law is
$\nabla_{\mu}\tilde T^{\mu\nu} = \tilde J_{\mu}F^{\mu\nu}$ 
and the trace anomaly is $T^{\mu}_{\quad \mu} = \frac{R}{24\pi}$
 (see also footnote 2) which have to be used instead of (\ref{new1}) and (\ref{new3}). Once again $\tilde{T}{^\mu}{_\nu}$ will not agree with our $T{^\mu}{_\nu}$ (\ref{37}). However, at asymptotic infinity, all components agree:
\begin{equation}
\tilde{T}{^\mu}{_\nu}(r\rightarrow\infty)=T{^\mu}{_\nu}(r\rightarrow\infty),
\label{ref4}
\end{equation}
leading to the identification of the Hawking flux with $\tilde{T}{^r}{_t}(r\rightarrow\infty)$.

   The equivalences (\ref{compare1},\ref{compare2},\ref{ref4}) reveal the internal consistency of our approach. They are based on two issues. First, in the asymptotic limit the chiral anomalies (\ref{3},\ref{6}) vanish and, secondly, the boundary conditions (\ref{a},\ref{b}) get identified with the Unruh state that is appropriate for discussing Hawking effect. It is important to note that, asymptotically,
 all the components, and not just the universal component that yields the flux, agree.  
 
%This happens because the chiral anomaly (\ref{3}) vanishes
%asymptotically thereby showing the consistency of our 
%approach. 

  In the Hartle-Hawking state, the conditions
$\tilde J_{u}(r \rightarrow r_{h}) =0$ and $\tilde J_{v}(r \rightarrow r_{h}) =0$ yield,
\begin{equation}
C_{1} = 0 \ ; \ C_{2} = \frac{e^2}{\pi}A_{t}(r_{h})\label{new9}
\end{equation}
so that, 
\begin{eqnarray}
\tilde J^{r}(r) &=& 0,\label{new10} \\
\tilde J^{t}(r) &=& \frac{e^2}{\pi f}(A_{t}(r) - A_{t}(r_{h}))~,\label{new11}
\end{eqnarray} 
Expectedly, there is no Hawking (charge) flux now. The 
above expressions, even at asymptotic infinity, do not
agree with our expressions (\ref{35},\ref{36}).

\subsection{Boulware vacuum }
Apart from the Unruh and Hartle-Hawking vacua there is another
vacuum named after Boulware \cite{boulware} which closely resembles the Minkowski vacuum asymptotically. 
In this vacuum, there is no radiation in the asymptotic future. In other words this implies $J^{r}$ and $T^{r}_{\quad t}$ given in (\ref{23}) and 
(\ref{25}) must vanish at $r\rightarrow \infty$ limit. Therefore, 
for the Boulware vacuum, we get
\begin{eqnarray}
 c + a &=& 0 \label{42}\\
 p &=& \frac{1}{4}z \label{43} 
\end{eqnarray} 
By substituting (\ref{42}) in (\ref{23}) and (\ref{24}) we have
\begin{eqnarray}
J^{r}(r)&=& \frac{e^2}{2\pi\sqrt{-g}}A_{t}(r)  \label{45}\\
J^{t}(r) &=& \frac{e^2}{2\pi f} A_{t}(r). \label{46}  
\end{eqnarray}
Similarly, by substituting (\ref{42}) and (\ref{43}) in equations (\ref{25} -\ref{26'}), we get   
\begin{eqnarray}
T^{r}_{\quad t} &=& \frac{e^2 A_{t}^2(r)}{4\pi\sqrt{-g}} + \frac{1}{\sqrt{-g}}N^{r}_{t}(r)\label{47}\\
T^{r}_{\quad r} &=& \frac{-e^2 A_{t}^2(r)}{4\pi f} - \frac{1}{f}N^{r}_{t}(r) + \frac{R}{96\pi} \label{48}\\
T^{t}_{\quad t} &=& \frac{e^2 A_{t}^2(r)}{4\pi f} +\frac{1}{f}N^{r}_{t}(r) + \frac{R}{96\pi}\label{49} 
\end{eqnarray}
Observe that there is no radiation in the asymptotic region in the Boulware
vacuum. Also, the trace anomaly (\ref{d}) is reproduced since this
is independent of the choice of quantum state.\\
Further, we note that, in the Kruskal coordinates, $J_{U}$ and $T_{UU}$
components of current and EM tensors diverge at the horizon. This can be 
seen by substituting equations (\ref{45}-\ref{46}) in (\ref{29}). Then
the expression for $J_{u}$ in Boulware vacuum becomes,
\begin{equation}
J_{u} = \frac{e^2}{2\pi}A_{t}(r) \label{f}
\end{equation}
while, by putting (\ref{47}-\ref{49}) in (\ref{31}), we 
obtain, for $T_{uu}$
\begin{equation}
T_{uu} = \frac{e^2 A_{t}^2(r)}{4\pi} + N^{r}_{t}(r)~. \label{g}
\end{equation}
Note that in the limit ($r\rightarrow r_{h}$) $J_{u}$ and $T_{uu}$ do
not vanish. Hence, in the Kruskal coordinates, the  current and EM tensor diverge. This is expected since the Boulware vacuum is not regular near the horizon. \\
%%%%%%%%%%%%%%%%%%%%%%%%%%%%%%%%%%%%%%%%%%%%%%%%%%%%%%%%%%%%%%%%%%%%%%%%%%%%%%%%%

\section{Examples}
We discuss two explicit examples where the Hawking flux and the 
complete expressions for the covariant current /EM tensor are
provided.\\

\subsection{Reissner-Nordstrom black hole} 

For this black hole, the metric in the $r-t$ sector is given by
\begin{equation}
ds^2 =  f(r)dt^2 - \frac{1}{f(r)}dr^2 \label{Reissner}
\end{equation}
with 
\begin{equation}
f(r) = 1 - \frac{2M}{r} + \frac{Q^2}{r^2} = \frac{(r-r_{+})(r-r_{-})}{r^2}
\end{equation} 
where $r_{\pm} = M\pm \sqrt{M^2 - Q^2}$ are the outer and inner horizons.  The gauge potential is given by $A_{t} = -\frac{Q}{r}$. Note that
in this case $\sqrt{-g} =1$.
We can easily write the expressions for various components of current and EM tensor for Unruh, Hartle-Hawking and Boulware vacua. 
As already discussed the results for Unruh and Hartle-Hawking vacua are
identical. In this case  we have from (\ref{35}) and (\ref{36})
\begin{eqnarray}
J^{r}(r)&=& \frac{e^2}{2\pi}[A_{t}(r) - A_{t}(r_{+})] \label{50}\\
J^{t}(r) &=& \frac{e^2 r^2}{2\pi (r-r_{+})(r-r_{-})} [A_{t}(r) - A_{t}(r_{
+})]\label{51}
\end{eqnarray}
The charge flux, obtained from the asymptotic limit of (\ref{50}) is,
\begin{equation}
J^{r}(r\rightarrow \infty) = -\frac{e^2}{2\pi}A_{t}(r_{+}) = \frac{e^2Q}{2\pi r_{+}} \label{h}
\end{equation}
reproducing the known result \cite{Isowilczek,shailesh}.\\
Similarly, the $r-t$ component of the covariant EM tensor    
 from (\ref{37}) is given by,
\begin{eqnarray}
T^{r}_{\quad t} &=& \frac{e^{2}}{4\pi}[A_{t}(r)-A_{t}(r_{+})]^2 + 
\frac{1}{192\pi}[2f(r)f''(r) - f'^{2}(r) + f'^{2}(r_{+})]  \label{i}
%T^{r}_{\quad r} &=& -\frac{r^2}{(r-r_{+})(r-r_{-})}\left[\frac{e^2}{4\pi}
%  [A_{t}(r)-A_{t}(r_{+})]^2 + \frac{1}{192\pi}[2f(r)f''(r) - f'^{2}(r) + f'^{2}(r_{+})]\right]\nonumber \\
%&& + \frac{f''}{96\pi}\label{j}\\
%T^{t}_{\quad t} &=&  -T^{r}_{\quad r}  + \frac{f''}{48\pi}.\label{k}
\end{eqnarray}
while, as before, the other components follow from (\ref{26},\ref{26'}).
As usual, the energy flux obtained from the asymptotic limit of 
(\ref{i}) yields,
\begin{eqnarray}
T^{r}_{\quad t }(r\rightarrow \infty) &=&  \frac{e^2Q^2}{4\pi r_{+}^2} +\frac{1}{192\pi}\left[\frac{2}{M+\sqrt{M^2-Q^2}}(M^2 + M\sqrt{M^2-Q^2} -Q^2)\right]^2 ~.\label{52}\\
\end{eqnarray}
This reproduces the usual expression of energy flux coming from
the Reissner-Nordstrom black hole \cite{Isowilczek,shailesh}.\\
For the Boulware vacuum, the expressions for current/EM tensors (\ref{45}-\ref{49}) are given by,
\begin{eqnarray}
J^{r} &=& \frac{e^2}{2\pi}A_{t}(r) \label{l}\\
J^{t} &=& \frac{e^2 r^2}{2\pi (r-r_{+})(r-r_{-})}A_{t}(r) \label{m}\\
T^{r}_{\quad t} &=& \frac{e^2}{4\pi}A^{2}_{t}(r) + \frac{1}{192\pi}[2ff'' -f'^{2}(r)] \label{n}\\
T^{r}_{\quad r} &=& - \frac{r^2}{(r-r_{+})(r-r_{-})}\left[\frac{e^2}{4\pi}A^{2}_{t}(r) + \frac{1}{192\pi}[2ff'' -f'^{2}(r)] \right] + \frac{f''}{96\pi}\label{o}\\
T^{t}_{\quad t} &=& -T^{r}_{\quad r} + \frac{f''}{48\pi}\label{p}
\end{eqnarray}
As we can observe, by taking the asymptotic limit of (\ref{l}) and (\ref{n}), there are no Hawking fluxes.  

\subsection{Garfinkle-Horowitz- Strominger (GHS) black hole}

GHS blackhole is a member of a family of solutions to low energy string theory \cite{strom1,strom2}. The  metric in the  $r-t$ sector of this black hole is given by \cite{sunshailesh, vega} 
\begin{equation}
ds^2 = f(r)dt^2 -\frac{1}{h(r)}dr^2 \label{GHS}
\end{equation} 
where 
\begin{eqnarray}
f(r) &=& (1-\frac{2Me^{\phi_{o}}}{r})(1-\frac{Q^2e^{3\pi_{o}}}{Mr})^{-1}\label{57}\\
h(r) &=& (1-\frac{2Me^{\phi_{o}}}{r})(1-\frac{Q^2e^{3\pi_{o}}}{Mr})~.\label{58}
\end{eqnarray}
with $\phi_{o}$ being the asymptotic constant value of the dilaton field.
We consider the case when $Q^2 < 2e^{-2\phi_{o}M^2}$ for which the above
metric describes a black hole with an event horizon \cite{sunshailesh,strom1, vega}, 
\begin{equation}
r_{h} = 2Me^{\phi_{o}}~.\label{59}
\end{equation}
Note that in this limit there is only one event horizon (\ref{59}) and 
the gauge fields will not play any role in the subsequent analysis. In other words we have only gravitational anomaly in the theory. Also, this is an
example with distinct $f(r), h(r)$ so that $\sqrt{-g}\ne1$. 
For this black hole  we can write the complete expressions for 
current/EM tensors for the Unruh (Hartle-Hawking)(\ref{37}) by
substituting the values for $f(r)$ (\ref{57}) and $h(r)$ (\ref{58}). For the  sake of simplicity, here we just give the asymptotic expression
for $T^{r}_{t}$
\begin{eqnarray}
T^{r}_{\quad t}(r\rightarrow \infty) &=& \frac{1}{192\pi}f'(r_{h})h'(r_{h}) = \frac{1}{768M^2e^{2\phi_{o}}}, \label{60}\\
\end{eqnarray} 
which gives the usual expression for energy flux from GHS black hole \cite{sunshailesh, vega}.\\
For the Boulware vacuum, substituting (\ref{57}) and (\ref{58}) in 
(\ref{47}) we note that there is no Hawking flux  in the  asymptotic region, as expected.

%%%%%%%%%%%%%%%%%%%%%%%%%%%%%%%%%%%%%%%%%%%%%%%%%%%%%%%%%%%%%
\section{Conclusions}
We have discussed in some details our method, briefly introduced in
\cite{Banerjee}, of computing the Hawking flux using covariant gauge
and gravitational anomalies. Contrary to earlier approaches, a split
of space into distinct regions (near to and away from horizon) using
step functions was avoided. This method is different from the anomaly
cancelling mechanism of \cite{Robwilczek,Isowilczek,Isoumtwilczek,shailesh}
although it uses identical (covariant) boundary conditions.
It reinforces the crucial role of these boundary conditions,
 the study of which has been the principal objective of this paper.

 In order to get a clean understanding of these boundary conditions
we first computed the explicit structures of the covariant current
$<J_{\mu}>$ and the covariant energy-momentum tensor $<T_{\mu\nu}>$
from the chiral (anomalous) effective action, appropriately modified by
adding a local counterterm \cite{shailesh2,Leut}. The chiral nature 
of these structures became more transparent by passing to the null coordinates. In these coordinates the contribution from the ingoing (left moving) modes was manifestly seen to vanish. The outgoing (right moving)
 modes involved arbitrary parameters which were fixed by imposing regularity conditions at the future horizon. 
No condition on the ingoing (left moving) modes was required
as these were absent as a result of chirality. These findings by themselves are new. They are also different
from the corresponding expressions for $<J_{\mu}>$, $<T_{\mu\nu}>$, 
obtained from the  standard nonanomalous (Polyakov) action,
satisfying $\nabla_{\mu}J^{\mu} = 0, \nabla_{\mu}T^{\mu\nu} = J_{\mu}F^{\mu\nu}$ and $T^{\mu}_{\quad \mu} = \frac{R}{24\pi}$, implying the absence of any gauge or gravitational (diffeomorphism) anomaly. Only the trace anomaly
is present. Details of the latter computation may be found in \cite{Isoumtwilczek,fabbri1}.\\

    We have then established a direct connection of these results with
the choice of the covariant boundary condition used in determining
the Hawking flux from chiral gauge and gravitational anomalies
\cite{Isowilczek,Isoumtwilczek,shailesh,shailesh2,Banerjee}. 
The relevant universal component ($J^{r}$ or $T^{r}_{\quad t}$) obtained by
solving the anomaly equation subject to the covariant boundary condition
(\ref{a},\ref{b}) agrees exactly with the result derived from
imposing regularity condition on the outgoing modes at the 
future horizon: namely, a free falling observer sees a finite amount of flux at outer horizon
indicating the possibility of Hawking radiation. Our findings, therefore,
 provide a clear justification of the covariant boundary condition.

 Finally, we put our computations in a proper perspective by comparing our findings with the standard implementation of the various vacua states on nonchiral expressions.
Specifically, we show that our results are compatible with the choice of Unruh vacuum for a nonchiral theory which eventually yields the Hawking flux.
 
%%%%%%%%%%%%%%%%%%%%%%%%%%%%%%%%%%%%%%%%%%%%%%%%%%%%%%%%%%%%%%%%%%
\section{Appendix}
Unlike  the case of vector theory, where the diffeomorphism invariance 
 is kept intact inspite of the presence of trace anomaly, the chiral theory
has both a diffeomorphism anomaly (gravitational anomaly) and a trace
anomaly. In $1+1$ dimensions it is possible to obtain a relation
between the coefficients of the diffeomorphism anomaly and the trace
anomaly by exploiting the chirality criterion. \\

  To see this let us write the general structure of the covariant Ward
identity,
\begin{equation}
\nabla_{\mu}T^{\mu}_{\quad \nu} = J_{\mu}F^{\mu}_{\quad \nu} +
N_{a}\bar \epsilon_{\nu\mu}\nabla^{\mu}R \label{A1} 
\end{equation}  
where $N_{a}$ is an undetermined normalisation. The functional form of the
anomaly follows on grounds of dimensionality, covariance and parity.
 Likewise, the structure of the covariant trace anomaly is written as ,
\begin{equation}
T^{\mu}_{\quad \mu} = N_{t}R \label{A2}  
\end{equation}
with $N_{t}$ being the normalisation.
In the null coordinates (\ref{27},\ref{28}) for $\nu = v$, the left hand side of (\ref{A1}) becomes
\begin{eqnarray}
\nabla_{\mu}T^{\mu}_{\quad v} &=& \nabla_{u}T^{u}_{\quad v} + 
\nabla_{v}T^{v}_{\quad v}\nonumber\\
&& = \nabla_{u}(g^{uv}T_{vv})+ \nabla_{v}(g^{uv}T_{uv})
 = \nabla_{v}(g^{uv}T_{uv}) \label{A3} 
\end{eqnarray}  
where we have used  the fact that for a chiral theory 
$T_{vv} =0$ (see equation \ref{32'}). 
Also, in  null coordinates, we have, 
\begin{equation}
T_{uv} = \frac{1}{2}(g_{uv}T^{v}_{\quad v} + g_{uv}T^{u}_{\quad u}) = \frac{g_{uv}}{2}T^{\mu}_{\quad \mu}= \frac{f}{4}T^{\mu}_{\quad \mu}~. \label{A4}
\end{equation}
By using (\ref{A2}), (\ref{A4}) and (\ref{A3}) we obtain,
\begin{equation}
\nabla_{\mu}T^{\mu}_{\quad v} = \frac{N_{t}}{2}\nabla_{v}R~. \label{A5}
\end{equation}
where we used $g^{uv}= \frac{2}{f}$ (\ref{nullmetric}).\\

 The right hand side of (\ref{A1}) for $\nu=v$, with the use of 
the chirality constraint $J_{v}=0$ (\ref{30}), yields
\begin{equation}
J_{\mu}F^{\mu}_{\quad v} +N_{a}\bar \epsilon_{v\mu}\nabla^{\mu}R = 
N_{a}\nabla_{v}R ~.\label{A6}
\end{equation}
Hence, by equating (\ref{A5}) and  (\ref{A6}) we find a relationship
between $N_{a}$ and $N_{t}$
\begin{equation}
N_{a} = \frac{N_{t}}{2}\label{A7} 
\end{equation}
which is compatible with (\ref{6}) and (\ref{d}) with $N_{a} = \frac{N_{t}}{2} = \frac{1}{96\pi}$. It is clear that chirality enforces both
the conformal and diffeomorphism anomalies. The trivial (anomaly free)
case $N_{a} = N_{t} =0$ is ruled out because, using general arguments
based on the unidirectional property of chirality, it is possible
to prove the existance of the diffeomorphism anomaly in
$1+1$ dimensions \cite{Fulling3}.
%%%%%%%%%%%%%%%%%%%%%%%%%%%%%%%%%%%%%%%%%%%%%%%%%%%%%%%%%%%%%%

%%%%%%%%%%%%%%%%%%%%%%%%%%%%%%%%%%%%%%%%%%%%%%%%%%%%%%%%%%%%%%%%%
\end{document}